\documentclass[twocolumn,prl,aps,amsmath,amssymb]{revtex4}

\usepackage{graphicx}
\usepackage{dcolumn}
\usepackage{bm}

\begin{document}
\title{Duality in the Color Flavor Locked Spectrum}
 \author{A. D. {\sc Jackson}}
 \email{jackson@alf.nbi.dk}
 \author{F. {\sc Sannino}}\email{francesco.sannino@nbi.dk}
 \affiliation{The Niels Bohr Institute \& {\rm NORDITA}, Blegdamsvej 17, DK-2100 Copenhagen
 \O,
Denmark}
\date{August 2003}

\begin{abstract}
We analyze the spectrum of the massive states for the color flavor
locked phase (CFL) of QCD. We show that the vector mesons have a
mass of the order of the color superconductive gap $\Delta$. We
also see that the excitations associated with the solitonic sector
of the CFL low energy theory have a mass proportional to
$F^2_{\pi}/\Delta$ and hence are expected to play no role for the
physics of the CFL phase for large chemical potential. Another
interesting point is that the product of the soliton mass and the
vector meson mass is independent of the gap. We interpret this
behavior as a form of electromagnetic duality in the sense of
Montonen and Olive. Our approach for determining the properties of
the massive states is non-perturbative in nature and can be
applied to any theory with multiple scales.
\end{abstract}

\maketitle

\label{uno} A color superconductivity phase is a reasonable
candidate for the state of strongly interacting matter for very
large quark chemical potential
\cite{{Barrois:1977xd},{Bar_79},{Bailin:1984bm},{Alford:1998zt},{Rapp:1998zu}}.
Many properties of such a state have been investigated for two and
three flavor QCD. In some cases these results rely heavily on
perturbation theory, which is applicable for very large chemical
potentials.

In this work we seek insight regarding the relevant energy scales
of various physical states of the color flavor locked phase (CFL),
such as the vector mesons and the solitons. Our results do not
support the naive expectation that all massive states are of the
order of the color superconductive gap, $\Delta$. Our strategy is
based on exploiting the significant information already contained
in the low--energy effective theory for the massless states. We
transfer this information to the massive states of the theory by
making use of the fact that higher derivative operators in the
low--energy effective theory for the lightest state can also be
induced when integrating out heavy fields. {}For the vector
mesons, this can be seen by considering a generic theory
containing vector mesons and Goldstone bosons. After integrating
out the vector mesons, the induced local effective Lagrangian
terms for the Goldstone bosons must match the local contact terms
from operator counting. We find that each derivative in the (CFL)
chiral expansion is replaced by a vector field $\rho_{\mu}$ as
follows
\begin{eqnarray}
\partial \rightarrow \frac{\Delta}{F_{\pi}}\rho\ . \end{eqnarray}
This relation allows us to deduce, among other things, that the
energy scale for the vector mesons is
\begin{eqnarray}
m_v \sim \Delta \ ,
\end{eqnarray}
where $m_v$ is the vector meson mass. Our result is in agreement
with the findings in \cite{{Casalbuoni:2000na},{Rho:2000ww}}. We
shall see that this also suggests that the KSRF relation holds in
the CFL phase.

{}In the solitonic sector, the CFL chiral Lagrangian
\cite{Hong:1999dk,{CG}} gives us the scaling behavior of the
coefficient of the Skyrme term and thus shows that the mass of the
soliton is of the order of
\begin{eqnarray}
M_{\rm soliton} \sim \frac{F^2_{\pi}}{\Delta} \ ,
\end{eqnarray}
which is contrary to naive expectations. This is suggestive of a
kind of duality between vector mesons and solitons in the same
spirit as the duality advocated some years ago by Montonen and
Olive for the $SU(2)$ Georgi-Glashow theory
\cite{Montonen:1977sn}. This duality becomes more apparent when
considering the product
\begin{eqnarray}
M_{\rm soliton} m_v \sim F^2_{\pi} \ ,
\end{eqnarray}
which is independent of the scale, $\Delta$. In the present case,
if the vector meson self-coupling is $\widetilde{g}$, we find that
the Skyrme coefficient, $e\sim \Delta/F_{\pi}$, can be identified
with $\widetilde{g}$.  Thus, the following relations hold:
\begin{eqnarray}
M_{\rm soliton} \propto \,\frac{F_{\pi}}{\widetilde{g}}  \quad
{\rm and}\quad m_{v}\propto \,\widetilde{g}\, F_{\pi} \ .
\end{eqnarray}
In this notation the electric-magnetic (i.e. vector meson-soliton)
duality is transparent. Since the topological Wess-Zumino term in
the CFL phase is identical to that in vacuum, we identify the
soliton with a physical state having the quantum numbers of the
nucleon. If quark-hadron continuity \cite{Schafer:1998ef} is
assumed, we expect that the product of the nucleon and vector
meson masses will scale like $F_{\pi}^2$ for any non-zero chemical
potential for three flavors. Interestingly, quark-hadron
continuity can be related to duality. Testing this relation can
also be understood as a quantitative check of quark-hadron
continuity. It is important to note that our results are tree
level results and that the resulting duality relation can be
affected by quantum corrections. Our results have direct
phenomenological consequences for the physics of compact stars
with a CFL phase.  While vector mesons are expected to play a
relevant role, solitons can safely be neglected for large values
of the quark chemical potential.

\section{The Lagrangian for CFL Goldstones}
When diquarks condense for the three flavor case, we have the
following symmetry breaking:
\begin{eqnarray}
\left[SU_c(3)\right] \times SU_L(3) \times SU_R(3) \times U_B(1)
\rightarrow SU_{c+L+R}(3) \ . \nonumber
\end{eqnarray}
The gauge group undergoes a dynamical Higgs mechanism, and nine
Goldstone bosons emerge. Neglecting the Goldstone mode associated
with the baryon number and quark masses (which will not be
important for our discussion at lowest order), the derivative
expansion of the effective Lagrangian describing the octect of
Goldstone bosons is \cite{Hong:1999dk,{CG}}:
\begin{eqnarray}
{\cal L}=\frac{F^2_{\pi}}{8}{\rm Tr}\left[\partial_{\mu}U
\partial^{\mu}U^{\dagger}\right]\equiv \frac{F^2_{\pi}}{2}
{\rm Tr}\left[p_{\mu}p^{\mu}\right] \ ,
\end{eqnarray}
with $p_{\mu}=\frac{i}{2} \left(\xi
\partial_{\mu}\xi^{\dagger} -
\xi^{\dagger}\partial_{\mu}\xi\right)$,  $U=\xi ^2$,
 $\xi=e^{i\frac{\phi}{F_{\pi}}}$ and $\phi$ is the octet of
Goldstone bosons. $U$ transforms linearly according to $g_L U
g_R^{\dagger}$  and $g_{L/R}\in SU_{L/R}(3)$ while $\xi$
transforms non-linearly:
\begin{eqnarray}
\xi \rightarrow g_L\,\xi \,K^{\dagger}\left(\phi,
g_L,g_R\right)\equiv K\left(\phi, g_L,g_R\right)\,\xi\,
g_R^{\dagger} \ .
\end{eqnarray}
This constraint implicitly defines the matrix, $K\left(\phi,
g_L,g_R\right)$.   Here, we wish to examine the CFL spectrum of
massive states using the technique of integrating in/out at the
level of the effective Lagrangian. $F_{\pi}$ is the Goldstone
boson decay constant.  It is a non-perturbative quantity whose
value is determined experimentally or by non-perturbative
techniques (e.g.\ lattice computation). For very large quark
chemical potential, $F_{\pi}$ can be estimated perturbatively. It
is found to be proportional to the Fermi momentum, $p_F\sim \mu$,
with $\mu$ the quark chemical potential \cite{Schafer:2003vz}.
Since a frame must be fixed in order to introduce a chemical
potential, spatial and temporal components of the effective
Lagrangians split. This point, however, is not relevant for the
validity of our results.

When going beyond the lowest-order term in derivatives, we need a
counting scheme.  For theories with only one relevant scale (such
as QCD at zero chemical potential), each derivative is suppressed
by a factor of $F_{\pi}$.  This is not the case for theories with
multiple scales. In the CFL phase, we have both $F_{\pi}$ and the
gap, $\Delta$, and the general form of the chiral expansion is
\cite{Schafer:2003vz}:
\begin{eqnarray}
L\sim F^2_{\pi}\Delta^2
\left(\frac{\vec{\partial}}{\Delta}\right)^{k}
\left(\frac{\partial_0}{\Delta}\right)^{l}
U^{m} {U^{\dagger}}^{n} \ .
\end{eqnarray}
Following \cite{Schafer:2003vz}, we distinguish between temporal
and spatial derivatives.  Chiral loops are suppressed by
powers of $p/4\pi F_{\pi}$, and higher-order contact terms are
suppressed by $p/\Delta$ where $p$ is the momentum. Thus, chiral
loops are parametrically small compared to contact terms when the
chemical potential is large.

There is also a topological term which is essential in order to
satisfy the t'Hooft anomaly conditions
\cite{S,{HSaS},{Sannino:2003pq}} at the effective Lagrangian
level.  It is important to note that respecting the t'Hooft
anomaly conditions is more than an academic exercise. In fact, it
requires that the form of the Wess-Zumino term is the same in
vacuum and at non-zero chemical potential. Its real importance
lies in the fact that it forbids a number of otherwise allowed
phases which cannot be ruled out given our rudimentary treatment
of the non-perturbative physics.  As an example, consider a phase
with massless protons and neutrons in three-color QCD with three
flavors.  In this case chiral symmetry does not break. This is a
reasonable realization of QCD for any chemical potential. However,
it does not satisfy the t'Hooft anomaly conditions and hence
cannot be considered.  Were it not for the t'Hooft anomaly
conditions, such a phase could compete with the CFL phase.

Gauging the Wess-Zumino term with to respect the electromagnetic
interactions yields the familiar $\pi^0\rightarrow 2\gamma$
anomalous decay. This term \cite{WZ} can be written compactly
using the language of differential forms. It is useful to
introduce the algebra-valued Maurer-Cartan one form $\alpha
=\alpha_{\mu}dx^{\mu}=\left( \partial _{\mu }U\right)
U^{-1}\,dx^{\mu }\equiv
\left( dU\right) U^{-1}$
which transforms only under the left $SU_{L}(3)$ flavor group. The
Wess-Zumino effective action is
\begin{eqnarray}
\Gamma _{WZ}\left[ U\right] =C\,\int_{M^{5}}{\rm Tr}\left[ \alpha
^{5}\right] \ .  \label{WZ}
\end{eqnarray}
The price which must be paid in order to make the action local is
that the spatial dimension must be augmented by one.  Hence, the
integral must be performed over a five-dimensional manifold whose
boundary ($M^{4}$) is ordinary Minkowski space. In
\cite{{Hong:1999dk},{S},Casalbuoni:2000jn} the constant $C$ has
been shown to be the same as that at zero density, i.e.
\begin{equation}
C=-i\frac{N_{c}}{240\pi ^{2}}\ , \label{coef}
\end{equation}
where $N_{c}$ is the number of colors (three in this case).  Due
to the topological nature of the Wess-Zumino term its
coefficient is a pure number.

\section{The vector mesons}
{}It is well known that massive states are relevant for low energy
dynamics.  Consider, for example, the role played by vector mesons
in pion-pion scattering \cite{Sannino:1995ik} in saturating the
unitarity bounds. More specifically, vector mesons play a relevant
role when describing the low energy phenomenology of QCD and may
also play a role also in the dynamics of compact stars with a CFL
core. In order to investigate the effects of such states, we need
to know their in--medium properties including their gaps and the
strength of their couplings to the CFL Goldstone bosons. Except
for the extra spontaneously broken $U(1)_B$ symmetry, the symmetry
properties of the CFL phase have much in common with those of zero
density phase of QCD.  This fact allows us to make some non
perturbative but reasonable estimates of vector mesons properties
in medium.  We have already presented the general form of the
chiral expansion in the CFL phase.  As will soon become clear, we
are now interested in the four--derivative (non--topological)
terms whose coefficients are proportional to
\begin{eqnarray}
\frac{F^2_{\pi}}{\Delta^2} \ .
\end{eqnarray}
This must be contrasted with the situation at zero chemical
potential, where the coefficient of the four--derivative term is
always a pure number before quantum corrections are taken into
account. In vacuum, the tree-level Lagrangian which simultaneously
describes vector mesons, Goldstone bosons, and their interactions
is:
\begin{eqnarray}
L&=&\frac{F^2_{\pi}}{2}{\rm
Tr}\left[p_{\mu}p^{\mu}\right]+\frac{m^2_{v}}{2} {\rm Tr}
\left[\left(\rho_{\mu} +
\frac{v_{\mu}}{\widetilde{g}}\right)^2\right] \nonumber \\
&-&\frac{1}{4}{\rm Tr}
\left[F_{\mu\nu}(\rho)F^{\mu\nu}(\rho)\right] \ ,
\end{eqnarray}
where $F_{\pi}\simeq 132$ Mev and $v_{\mu}$ is the one form
$v_{\mu}=\frac{i}{2} \left(\xi \partial_{\mu}\xi^{\dagger} +
\xi^{\dagger}\partial_{\mu}\xi\right)$
with $U=\xi^2$ and $F_{\mu\nu}(\rho)=\partial_{\mu}\rho_{\nu} -
\partial_{\nu}\rho_{\mu} + i \widetilde{g}\,
[\rho_{\mu},\rho_{\nu}]$. At tree level this Lagrangian agrees
with the hidden local symmetry results \cite{Bando:1987br}.

When the vector mesons are very heavy with respect to relevant momenta,
they can be integrated out. This results in the field contraint:
\begin{eqnarray}
\rho_{\mu} = -\frac{v_{\mu}}{\widetilde{g}} \ .
\end{eqnarray}
Substitution of this relation in the vector meson kinetic term
(i.e., the replacement of $F_{\mu\nu}(\rho)$ by $F_{\mu\nu}(v)$)
gives the following four derivative operator with two time
derivatives and two space derivatives \cite{Schechter:1999hg}:
\begin{eqnarray}  \frac{1}{64\,\widetilde{g}^2}{\rm Tr}\left[[\alpha_{\mu}
,\alpha_{\nu}]^2\right] \ .\label{4d}\end{eqnarray} The
coefficient is proportional to $1/\widetilde{g}^2$. It is also
relevant to note that since we are describing physical fields we
have considered canonically normalized fields and kinetic terms.
This Lagrangian can also be applied to the CFL case. In the
vacuum, $\widetilde{g}$ is a number of order one independent of
the scale at tree level. This is no longer the case in the CFL
phase. Here, by comparing the coefficient of the four--derivative
operator in eq.~(\ref{4d}) obtained after having integrated out
the vector meson with the coefficient of the same operator in the
CFL chiral perturbation theory we determine the following scaling
behavior of $\widetilde{g}$:
\begin{eqnarray}
\widetilde{g} \propto \frac{\Delta}{F_{\pi}} \ . \label{first}
\end{eqnarray}
By expanding the effective Lagrangian with the respect to the
Goldstone boson fields, one sees that $\widetilde{g}$ is also
connected to the vector meson coupling to two pions, $g_{\rho \pi
\pi}$, through the relation
\begin{eqnarray}
g_{\rho\pi \pi} =
\frac{m^2_{v}}{\widetilde{g}F^2_{\pi}} \ .
\end{eqnarray}
In vacuum $g_{\rho\pi \pi } \simeq 8.56$ and $\widetilde{g} \simeq
3.96$ are quantities of order one. Since $v_{\mu}$ is essentially
a single derivative, the scaling behavior of $\widetilde{g}$
allows us to conclude that each derivative term is equivalent to
$\widetilde{g}\,\rho_{\mu}$ with respect to the chiral expansion.
For example, dropping the dimensionless field $U$, the operator
with two derivatives becomes a mass operator for the vector meson
\begin{eqnarray}
F^2_{\pi} \partial^2_{\mu} \rightarrow F_{\pi}^2 \,
\widetilde{g}^2\, \rho^2_{\mu} \sim \Delta^2 \rho^2_{\mu}\ .
\end{eqnarray}
This demonstrates that the vector meson mass gap is proportional
to the color superconducting gap. This non-perturbative result is
relevant for phenomenological applications. It is interesting to
note that our simple counting argument agrees with the underlying
QCD perturbative computations of Ref.~\cite{Casalbuoni:2000na} and
also with recent results of Ref.~\cite{Rho:2000ww}. However, our
approach is more general since it does not rely on any underlying
perturbation theory. It can be applied to theories with multiple
scales for which the counting of the Goldstone modes is known.
Since $m^2_v \sim \Delta^2$, we find that $g_{\rho\pi\pi}$ scales
with $\widetilde{g}$ suggesting that the KSRF relation is a good
approximation also in the CFL phase of QCD.

\section{CFL-Solitons}
The low energy effective theory supports solitonic excitations
which can be identified with the baryonic sector of the theory at
non-zero chemical potential. In order to obtain classically stable
configurations, it is necessary to include at least a
four--derivative term (containing two temporal derivatives) in
addition to the usual two--derivative term. Such a term is the
Skyrme term:
\begin{eqnarray}
L^{\rm skyrme}=\frac{1}{32\,e^2}{\rm Tr}\left[[\alpha_{\mu}
,\alpha_{\nu}]^2\right] \ .
\end{eqnarray}
Since this is a fourth--order term in derivatives not associated
with the topological term we have:
\begin{eqnarray}
e \sim \frac{\Delta}{ F _{\pi}} \ .
\end{eqnarray}
This term is the same as that which emerges after integrating out
the vector mesons (see eq.~(\ref{4d})), and one concludes that $e
=\sqrt{2}\,\widetilde{g}$ \cite{Schechter:1999hg}. The simplest
complete action supporting solitonic excitations is:
\begin{eqnarray}
\int\,d^4x\left[\frac{F^2_{\pi}}{2}{\rm
Tr}\left[p_{\mu}p^{\mu}\right] + L^{\rm skyrme}\right] +
\Gamma_{WZ} \ .
\end{eqnarray}
The Wess-Zumino term in eq.~(\ref{WZ}) guarantees the correct
quantization of the soliton as a spin $1/2$ object. Here we
neglect the breaking of Lorentz symmetries, irrelevant to our
discussion. The Euler-Lagrangian equations of motion for the
classical, time independent, chiral field $U_0(\bf{r})$ are highly
non-linear partial differential equations. To simplify these
equations Skyrme adopted the hedgehog {\it ansatz} which, suitably
generalized for the three flavor case, reads
\cite{Schechter:1999hg}: \begin{eqnarray} U_0(\bf{r}) =\left(%
\begin{array}{cc}
   e^{i {\vec{\tau}}\cdot \hat{r} F(r)}& 0 \\
  0& 1 \\
\end{array}%
\right)\ ,
\end{eqnarray}
where $\vec{\tau}$ represents the Pauli matrices and the radial
function $F(r)$ is called the chiral angle. The {\it ansatz} is
supplemented with the boundary conditions $F(\infty)=0$ and
$F(0)=0$ which guarantee that the configuration posseses unit
baryon number. After substituting the {\it ansatz} in the action
one finds that the classical solitonic mass is, up to a numerical
factor:
\begin{eqnarray}
M_{\rm soliton} \propto \frac{F_{\pi}}{e} \sim
\frac{F^2_{\pi}}{\Delta}\ ,
\end{eqnarray}
and the isoscalar radius, $\langle r^2\rangle_{I=0}\sim
1/({F^2_{\pi}\,e^2})\sim 1/\Delta^2$. Interestingly, due to the
non perturbative nature of the soliton, its mass turns to be dual
to the vector meson mass. It is also clear that although the
vector mesons and the solitons have dual masses, they describe two
very distinct types of states. The present duality is very similar
to the one argued in \cite{Montonen:1977sn}. Indeed, after
introducing the collective coordinate quantization, the soliton
(due to the Wess-Zumino term) describes baryonic states of
half-integer spin while the vectors are spin one mesons. Here, the
dual nature of the soliton with respect to the vector meson is
enhanced by the fact that, in the CFL state, $\widetilde{g}\sim
\Delta/F_{\pi}$ is expected to be substantially reduced with
respect to its value in vacuum. Once the soliton is identified
with the nucleon (whose density--dependent mass is denoted with
$M_N(\mu)$) and assuming quark-hadron continuity, we predict the
following relation to be independent of the matter density:
\begin{eqnarray}
\frac{M_{N}(\mu)\,m_{v}(\mu)}{(2\pi F_{\pi}(\mu))^2} =
\frac{M_{N}(0)\,m_{v}(0)}{(2\pi F_{\pi}(0))^2}\sim 1.05 \ .
\end{eqnarray}
In this way, we can relate duality to quark-hadron continuity.

\section{Conclusions}
We have shown that the vector mesons in the CFL phase have masses
of the order of the color superconductive gap, $\Delta$. On the
other hand the solitons have masses proportional to
$F^2_{\pi}/\Delta$ and hence should play no role for the physics
of the CFL phase at large chemical potential. We have noted that
the product of the soliton mass and the vector meson mass is
independent of the gap. This behavior reflects a form of
electromagnetic duality in the sense of Montonen and Olive
\cite{Montonen:1977sn}. Combining duality and quark-hadron
continuity we have predicted that the nucleon mass times the
vector meson mass scales as the square of the pion decay constant
at any non zero chemical potential. In the presence of two or more
scales provided by the underlying theory the spectrum of massive
states shows very different behaviors which cannot be obtained by
assuming a naive dimensional analysis.

It is a pleasure to thank R. Casalbuoni for discussions and J.
Schechter for careful reading of the manuscript. The work of F.S.
is supported by the Marie--Curie fellowship under contract
MCFI-2001-00181.

\end{document}